\documentclass[usenatbib,useAMS]{mn2e}
\usepackage{graphics}
\usepackage{psfrag}
\usepackage{xspace}
\usepackage{amstext}
\usepackage{amsmath}
\usepackage{amssymb}
\usepackage{natbib}
\usepackage{aas_macros}
\usepackage{ulem}
\usepackage{graphicx}
\usepackage{longtable}
\usepackage{rotating}
\usepackage{float}

\newcommand{\FigDir}[1]{#1}

\newcommand{\umin}{\ensuremath{\,u_{\rm min}}\xspace}
\newcommand{\Dl}{\ensuremath{D_{\rm l}}\xspace}

\newcommand{\Ds}{\ensuremath{D_{\rm s}}\xspace}
\newcommand{\tE}{\ensuremath{t_{\rm E}}\xspace}
\newcommand{\tN}{\ensuremath{t_{\rm N}}\xspace}

\newcommand{\Ml}{\ensuremath{M_{\rm l}}\xspace}

\newcommand{\tzero}{\ensuremath{t_{0}}\xspace}
\newcommand{\rE}{\ensuremath{R_{\rm E}}\xspace}

\newcommand{\Amax}{\ensuremath{A_{\rm max}}\xspace}
\newcommand{\vperp}{\ensuremath{v_{\perp}}\xspace}
\newcommand{\rs}{\ensuremath{r_{\rm s}}\xspace}

\newcommand{\Msun}{\ensuremath{M_{\odot}}\xspace}

\newcommand{\Rsun}{\ensuremath{R_{\odot}}\xspace}

\newcommand{\kpc}{\ensuremath{\rm kpc}\xspace}

\newcommand{\kmps}{\ensuremath{\rm km\,s^{-1}}\xspace}
\newcommand{\au}{{\rm AU}\xspace}

\newcommand{\Nul}{\ensuremath{N_{\rm \mu l}}\xspace}
\newcommand{\Pcb}{\ensuremath{P_{\rm cb}}\xspace}
\newcommand{\Pbl}{\ensuremath{P_{\rm bl}}\xspace}
\newcommand{\Pcc}{\ensuremath{P_{\rm cc}}\xspace}
\newcommand{\Pgv}{\ensuremath{P_{\rm gv}}\xspace}
\newcommand{\thc}{\ensuremath{\theta_{\rm c}}\xspace}

\newlength{\voff}
\begin{document}

\title[Microlensing of close binary stars]{Microlensing of close binary stars}

\author[Rattenbury, N.~J.]
{Nicholas J. Rattenbury$^{1}$\thanks{e-mail: (nicholas.rattenbury)@manchester.ac.uk}\\
$^1$ Jodrell Bank Centre for Astrophysics, Alan Turing Building, The University of Manchester, Manchester, M13 9PL, UK \\
}
\date{Accepted ........
      Received .......;
      in original form ......}

\pubyear{2007}

\maketitle
\begin{abstract}
The gravity due to a multiple-mass system has a remarkable gravitational effect: the extreme magnification of background light sources along extended so-called caustic lines. This property has been the channel for some remarkable astrophysical discoveries over the past decade, including the detection and characterisation of extra-solar planets, the routine analysis of limb-darkening,  and, in one case, limits set on the apparent shape of a star several kiloparsec distant. In this paper we investigate the properties of the microlensing of close binary star systems. We show that in some cases it is possible to detect flux from the Roche lobes of close binary stars. Such observations could constrain models of close binary stellar systems.
\end{abstract}


\section{Introduction}
\label{sec:intro}
Microlensing is the the time-dependent amplification of light from background stars by the gravitational field of massive object passing close to the observer-source line of sight \citep{1964PhRv..133..835L}. Initially proposed as a means for discovering the mass fraction of compact dark matter in the Galactic halo \citep{1986ApJ...304....1P}, the phenomenon of microlensing has been remarkably successful as an astrophysical tool for a variety of purposes. The technique has most notably proved a fruitful channel for the detection of extra-solar planets orbiting the lens star, complementing the radial velocity and transit planet detection techniques (see e.g. \citealt{2008arXiv0803.4324G}). Microlensing events can also yield information on the source object: the finite size of microlensed stars is routinely estimated, along with models of source star limb-darkening (e.g. \citealt{2001ApJ...549..759A,2003A&A...411L.493A,2003ApJ...596.1305F}). In one case, the projected source star shape was constrained by microlensing observations with an effective resolution of $0.04$ microarcsec \citep{2005A&A...439..645R}. The most spectacular results from microlensing occur as a result of the gravitational field of a binary (or multiple) lens object being able to highly amplify small sections of the source object. In this paper we investigate how the phenomenon of binary lens microlensing might be able to allow the detection and characterisation of close binary systems in the Galactic bulge, including the possibility of resolving extended, non-circular features of close- to contact binary systems. In Section~\ref{sec:microlensing} we describe the microlensing phenomenon and in Section~\ref{sec:close} we show example microlensing light curves arising from close binary systems, including contact binary systems showing non-spherical atmospheres. Section~\ref{sec:discussion} contains a discussion on these results, likelihood of observation, and suggestions for ensuring that these effects may be observed.

\section{Microlensing theory and practice}
\label{sec:microlensing}
The linear and temporal scales for microlensing are most conveniently expressed in units of the Einstein ring radius in the lens plane, \rE,  and event time \tE:

\begin{eqnarray}
\rE  &=& \left[\frac{4G\Ml}{c^2}\frac{\Dl(\Ds - \Dl)}{\Ds}\right]^{1/2} \rm{m} \notag \\ 
 &=& 4.42 \sqrt{\frac{\Ml}{0.3\Msun}}\sqrt{\frac{\Ds}{8\kpc}}\sqrt{x(1-x)} \hspace{25pt} \au \\
\tE &=& \frac{\rE}{\vperp} \notag \\
 &= &34.78\sqrt{\frac{\Ml}{0.3\Msun}}\sqrt{\frac{\Ds}{8\kpc}}\sqrt{x(1-x)}  \hspace{25pt} \rm{days}
\end{eqnarray}

\noindent where \Ml is the mass of the lens object, \Dl and \Ds are the distances to the lens and source objects respectively\footnote{All quantities are in SI units} and $x = \Dl / \Ds$. For Galactic microlensing, the lens transverse velocity, $\vperp = 220 \kmps$.

For a single mass acting as the lens, the amplification of the source star smoothly increases to the point where the source and lens systems are maximally co-aligned, with minimum impact parameter \umin, and then decreases as the lens and source systems move out of alignment. For a point-like source object, the maximum amplification is $\Amax = \umin^{-1}$. For a lens system with two or more mass elements, the light curve can change abruptly, showing sharp changes in amplification. Source plane loci for a multi-element lens where the background source star is strongly amplified by the gravitational potential of the lens system take the form of closed curves called caustics. Caustic-crossing events have the potential to allow strong constraints to be placed on the lens and source systems. The phenomenon of multiple-mass lensing is highly non-linear, a property that presents challenges and opportunities to modellers.

\section{Microlensing of close binary systems}
\label{sec:close}
\begin{figure}
\psfrag{umin}[][]{\hspace{-15pt}\umin}
\psfrag{beta}[][]{\hspace{-15pt}$\beta$}
\psfrag{xlab1}[][]{$\rE$}
\psfrag{ylab1}[][]{$\rE$}
\psfrag{d}[][]{$d$}
\psfrag{d}[][]{d}
\begin{center}
\hspace{-1cm}
\centering\includegraphics[width=1.0\hsize]{\FigDir{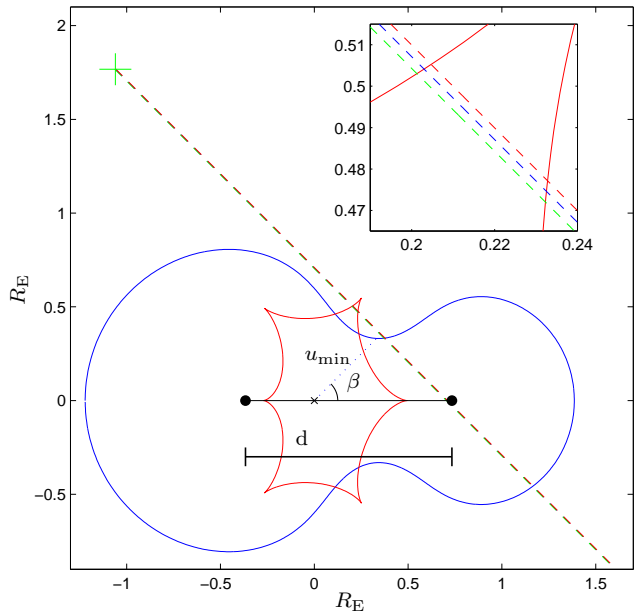}}
\end{center}
\caption{Example caustic crossing event. The normal to the source star track makes an angle $\beta$ to the line connecting the two lens masses, separated by a distance, $d$. \umin is the minimum source impact parameter to the lens system centre-of-mass. The caustic lines of the system are indicated by the curved red lines. The source travels along the source track from the upper left, crossing the caustic lines twice, resulting in large changes in source amplification, see Figure~\ref{fig:geometry2}. The inset shows the caustic crossing region with two extra lines denoting the extent of the finite source star size.}
\label{fig:geometry1}
\end{figure}

\begin{figure}
\psfrag{xlab2}[][]{$\tN$}
\psfrag{ylab2}[][]{Amplification}
\begin{center}
\hspace{-1cm}
\centering\includegraphics[width=1.0\hsize]{\FigDir{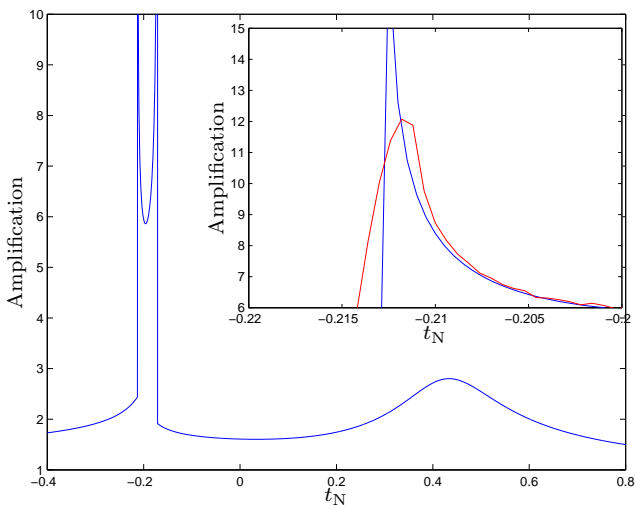}}
\end{center}
\caption{Light curves corresponding to the caustic crossing system illustrated in Figure~\ref{fig:geometry1}. The blue curve shows the light curve assuming a point-source, the red curve assuming a source star size $\rs = \Rsun$. For this and following light curve plots we show source system amplification versus the normalised time co-ordinate, $\tN = (t_{\rm JD} - \tzero)/\tE$, where $t_{\rm JD}$ is measured in days, \tzero is the epoch of closest projected source approach to the lens system centre-of-mass and $\tE=35$ days.}
\label{fig:geometry2}
\end{figure}

In this paper, we restrict the discussion to binary lens systems for reasons of simplicity. Figure~\ref{fig:geometry1} shows the caustic curves of a binary lens system and the path of a single example source star. The resulting light curve for both a point and finite-sized source star is shown in Figure~\ref{fig:geometry2}. The caustic crossing features are clearly visible in the light curve. A complete discussion of binary lens microlensing is not required here; it is sufficient to note that the passage of a source star across a caustic line results in the extreme amplification of the source star at the position of the caustic. The caustic pattern of a lens system comprised of two masses, $m_{1},m_{2}$ depends on the lens mass ratio $q_{\rm l} = m_{2}/m_{1}$ and the separation of the masses $d$. The resulting light curve depends on the passage of the source star  across the magnification profile produced by the lens system, specified by \umin, \tE and the angle of the source star track to the projected lens separation axis. For a more complete discussion on binary lens microlensing, see \citet{2006MPLA...21..919R} and references therein.

\begin{table*}
\begin{tabular}{ccccccccc}
\hline
Name & Period & Sp1+Sp2  & $q_{\rm s}$ & $R_{1}$ & $R_{2}$ & $a$ & $J_{1}/J_{2}$ & $i$\\
&(days) & & $m_{2}/m_{1}$ & (\Rsun) & (\Rsun) & (\Rsun) & & (deg)\\
\hline
TW And & 4.12  & FV+KIV & 0.19 & 2.05 & 3.20 & 13.6 & 10.4 & 86.9\\
HH Car & 3.23  & 8V+BIII  & 0.82 & 6.1 & 10.7 & 28.9 & 1.2 & 81.5\\
RZ Sct & 15.19 & B3Ib+F5IV & 0.21 & 15 & 15.9 & 62.46 & 15.3 & 82.5\\
V356 Sgr & 8.90 & B3V+A2II  & 0.39 & 7.4 & 14 & 46.26 & 5.2 & 82.6\\
\hline
\end{tabular}
\caption{Parameters of the close binary stars used as source systems in this work. From  \citet{2004yCat.5115....0S}. Listed are each system's orbital period, component spectral types, mass ratio $q_{\rm s}$, component stellar radii and orbital radius in units of solar radius, the ratio of surface brightness $J_{1}/J_{2}$ and orbital inclination.}
\label{tab:eclipdata}
\end{table*}

\subsection{Close binary stars as microlensing sources}
\label{sec:closebinary}
\begin{figure}
\psfrag{ylabel}[][]{Amplification}
\psfrag{xlabel}[][]{$\tN$}
\begin{center}
\hspace{-1cm}
\centering\includegraphics[width=1.0\hsize]{\FigDir{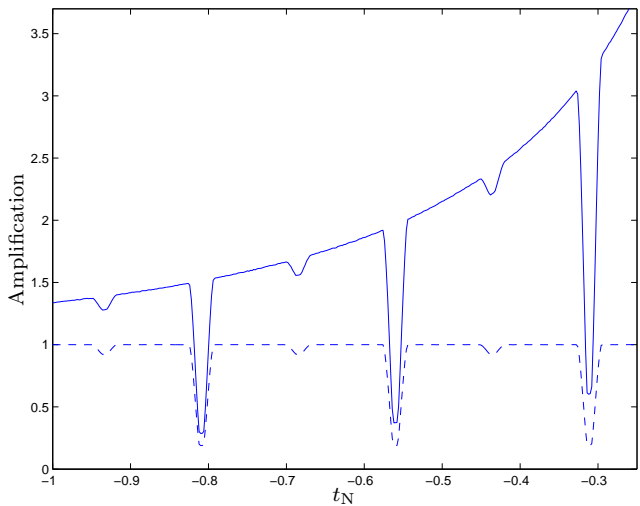}}
\end{center}
\caption{Part of an example single lens microlensing light curve assuming the source system is TW Andromedae. Lens and source distances are \Dl = 6\,\kpc and \Ds = 8\,\kpc respectively, \tE = 35 days, \Ml = 0.3\Msun and $\umin=0.1$. The unlensed light curve (dashed line) is also shown. }
\label{fig:TWAnd}
\end{figure}

We consider now the effect of a close binary acting as the source in a microlensing event. We use the catalogue of eclipsing binary stars of \citet{2004yCat.5115....0S}. Figure~\ref{fig:TWAnd} shows part of the light curve of the eclipsing binary TW Andromedae during a single lens microlensing event. \citet{1997ApJ...480..196H} showed that the Einstein ring radius, \rE, could be determined from binary-source single lens microlensing events. Single mass lens systems will not be considered further here except for the case of events with very high maximum amplification, see Section~\ref{sec:single}. 

\begin{figure}
\psfrag{ylab}[][]{Amplification}
\psfrag{xlab}[][]{$\tN$}
\begin{center}
\hspace{-1cm}
\centering\includegraphics[width=1.0\hsize]{\FigDir{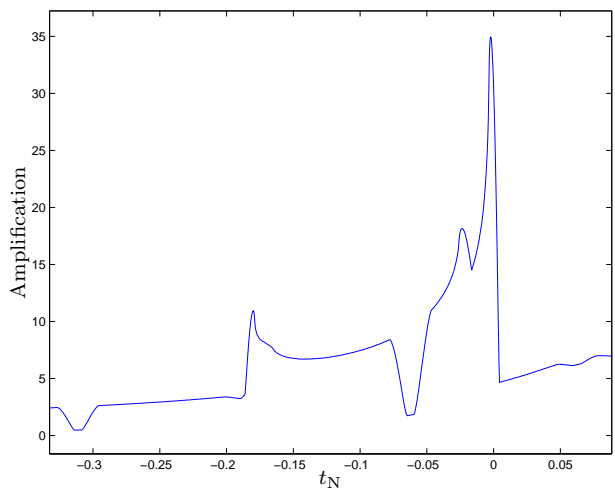}}
\end{center}
\caption{ Light curve of TW And arising from a binary lens with mass ratio $q_{\rm l}=0.11$, $\umin=0.1$, $d=0.95$, $\Ml = 0.3\Msun$, $\Dl=6\,\kpc$, $\Ds=8\,\kpc$.  }
\label{fig:TW_And_curve}
\end{figure}

\begin{figure}
\psfrag{xlab1}[][]{$\rE$}
\psfrag{ylab1}[][]{$\rE$}
\begin{center}
\hspace{-1cm}
\centering\includegraphics[width=1.0\hsize]{\FigDir{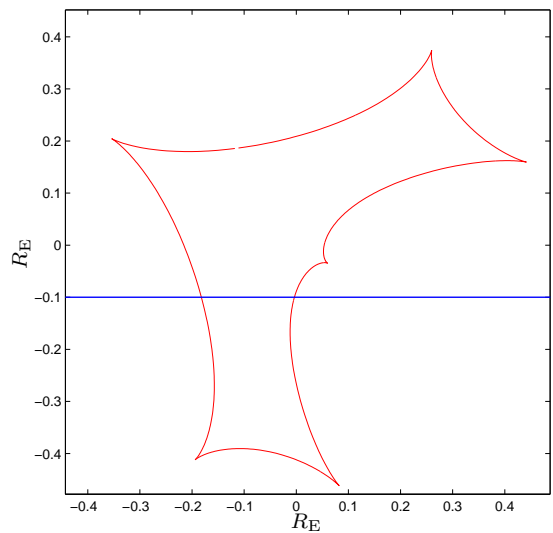}}
\end{center}
\caption{Caustic lines and source system track for the light curve shown in Figure~\ref{fig:TW_And_curve}. The centre-of-mass of the binary source system moves along the source system  track shown in blue from left to right.}
\label{fig:TW_And_caustic}
\end{figure}

\begin{figure}
\psfrag{ylab}[][]{Amplification}
\psfrag{xlab}[][]{$\tN$}
\begin{center}
\hspace{-1cm}
\centering\includegraphics[width=1.0\hsize]{\FigDir{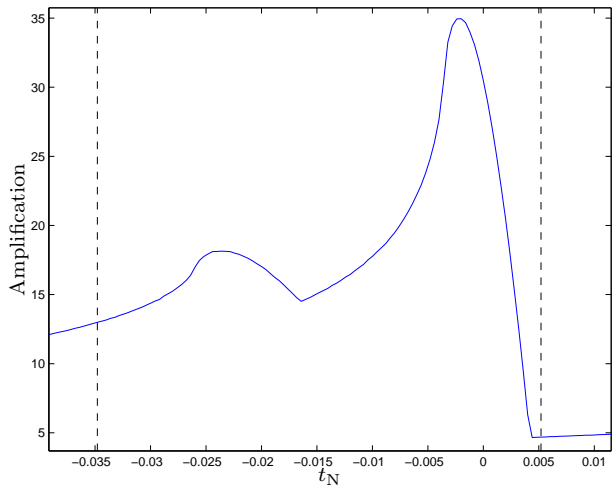}}
\end{center}
\caption{Detail of light curve shown in Figure~\ref{fig:TW_And_curve} showing the ``repeated'' caustic characteristic of a close binary source. The vertical dashed lines correspond to the source star system locations and orbital phase shown in Figure~\ref{fig:TW_And_caustic_zoom}.}
\label{fig:TW_And_curve_zoom}
\end{figure}

\begin{figure}
\begin{center}
\psfrag{xlab1}[][]{$\rE$}
\psfrag{ylab1}[][]{$\rE$}
\hspace{-1cm}
\centering\includegraphics[width=1.0\hsize]{\FigDir{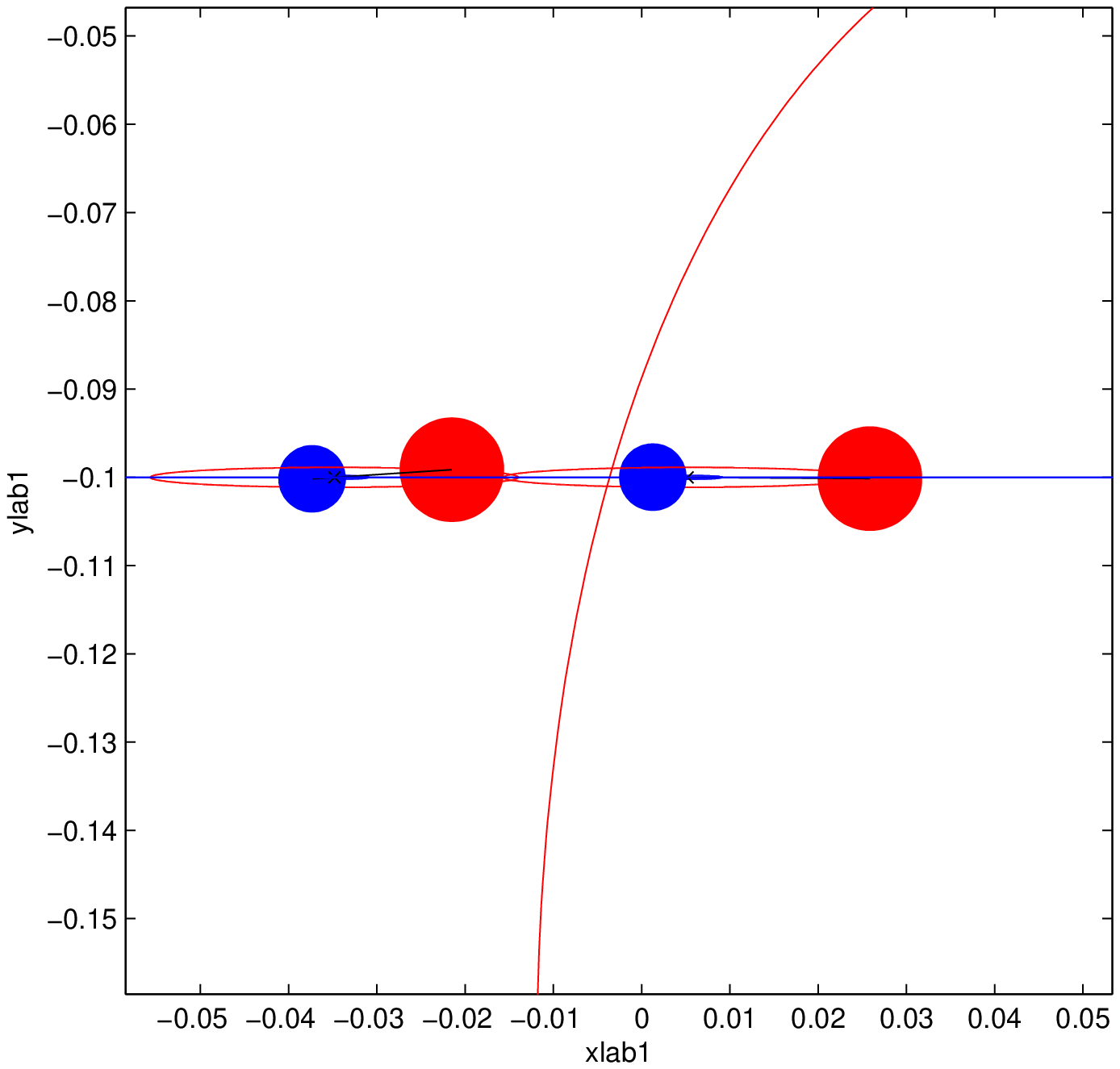}}
\end{center}
\caption{Caustic crossing detail corresponding to the light curve shown in Figure~\ref{fig:TW_And_curve_zoom} with the source star system location and orbital phase shown for the two times indicated by vertical dashed lines in Figure~\ref{fig:TW_And_curve_zoom}. The centre-of-mass of the binary source system moves along the source system  track shown in blue from left to right.}
\label{fig:TW_And_caustic_zoom}
\end{figure}

In contrast, Figure~\ref{fig:TW_And_curve} shows the light curve arising from the same close binary system amplified by a binary star lens. The source parameters for TW And are listed in Table~\ref{tab:eclipdata} and the microlensing parameters are as listed in the caption of Figure~\ref{fig:TW_And_curve}. The source star track across the caustic lines arising from the binary lens system is shown in Figure~\ref{fig:TW_And_caustic}. Figures~\ref{fig:TW_And_curve_zoom} and \ref{fig:TW_And_caustic_zoom} show light curve details at one of the caustic crossings and the corresponding source star system interaction with the lens system caustic. The source system is shown at two orbital phases to illustrate the relative source orbital duration relative to the microlensing event time. The complicated light curve shown in Figure~\ref{fig:TW_And_curve} is interpreted as a combination of source eclipses and caustic crossing features. The light curves shown in this work where generated using a modified version of the inverse ray-shooting method described in \citet{2002MNRAS.335..159R}.

\begin{figure}
\psfrag{ylab}[][]{Amplification}
\psfrag{xlab}[][]{$\tN$}
\begin{center}
\hspace{-1cm}
\centering\includegraphics[width=1.0\hsize]{\FigDir{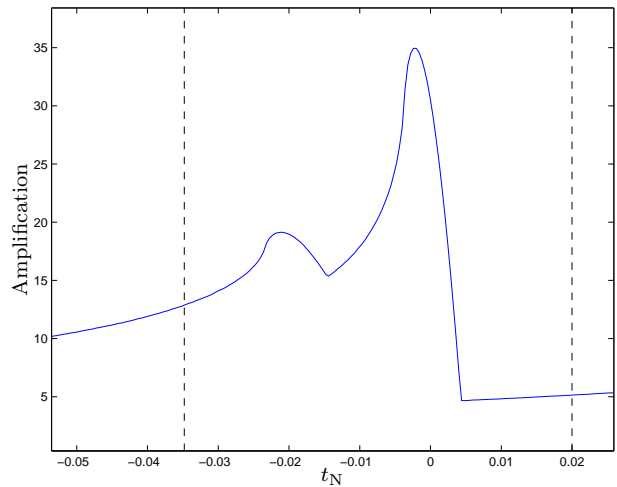}}
\end{center}
\caption{Detail of light curve for source system TW And assuming inclination angle $i=0$. The complete light curve shows no sign of binary star eclipses, but still displays  the ``repeated'' caustic characteristic of a close binary source. The vertical dashed lines correspond to the source star system locations and orbital phases shown in Figure~\ref{fig:TW_And_caustic_faceon}. }
\label{fig:TW_And_curve_faceon}
\end{figure}

\begin{figure}
\begin{center}
\psfrag{xlab1}[][]{$\rE$}
\psfrag{ylab1}[][]{$\rE$}
\hspace{-1cm}
\centering\includegraphics[width=1.0\hsize]{\FigDir{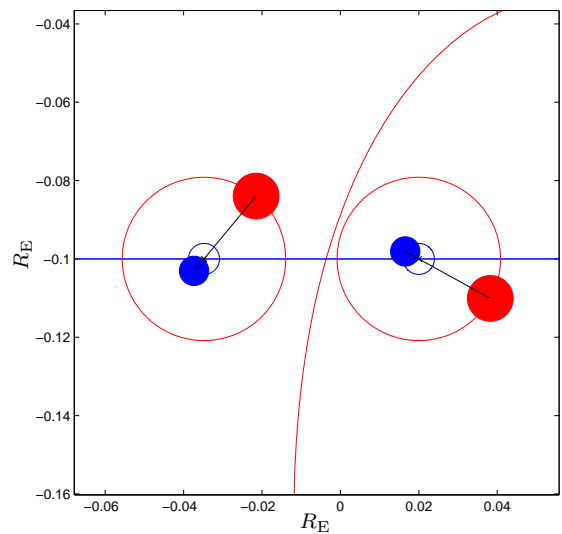}}
\end{center}
\caption{Caustic crossing detail of the light curve shown in Figure~\ref{fig:TW_And_curve_faceon} with the source star system location and orbital phase shown for the two times indicated by vertical dashed lines in Figure~\ref{fig:TW_And_curve_faceon}. The centre-of-mass of the binary source system moves along the source system  track shown in blue from left to right.}
\label{fig:TW_And_caustic_faceon}
\end{figure}

Of potentially more interest is the situation illustrated in Figures~\ref{fig:TW_And_curve_faceon} and \ref{fig:TW_And_caustic_faceon}, where the same source system, TW And, is shown amplified by the same lens system but where the inclination angle has been changed to give an non-eclipsing close binary system. The presence of a second source star is easily observed by the ``repeated'' caustic crossings seen in Figure~\ref{fig:TW_And_curve_faceon}. 

\begin{figure*}
\psfrag{xlab}[][]{\tN}
\psfrag{ylab}[][]{Amplification}
\psfrag{xlab1}[][]{$\rE$}
\psfrag{ylab1}[][]{$\rE$}
\psfrag{label1}[][]{}
\psfrag{label2}[][]{}
\psfrag{label3}[][]{}
\psfrag{label4}[][]{}
\psfrag{label5}[][]{}
\psfrag{label6}[][]{}
\centering\includegraphics[width=0.45\hsize]{\FigDir{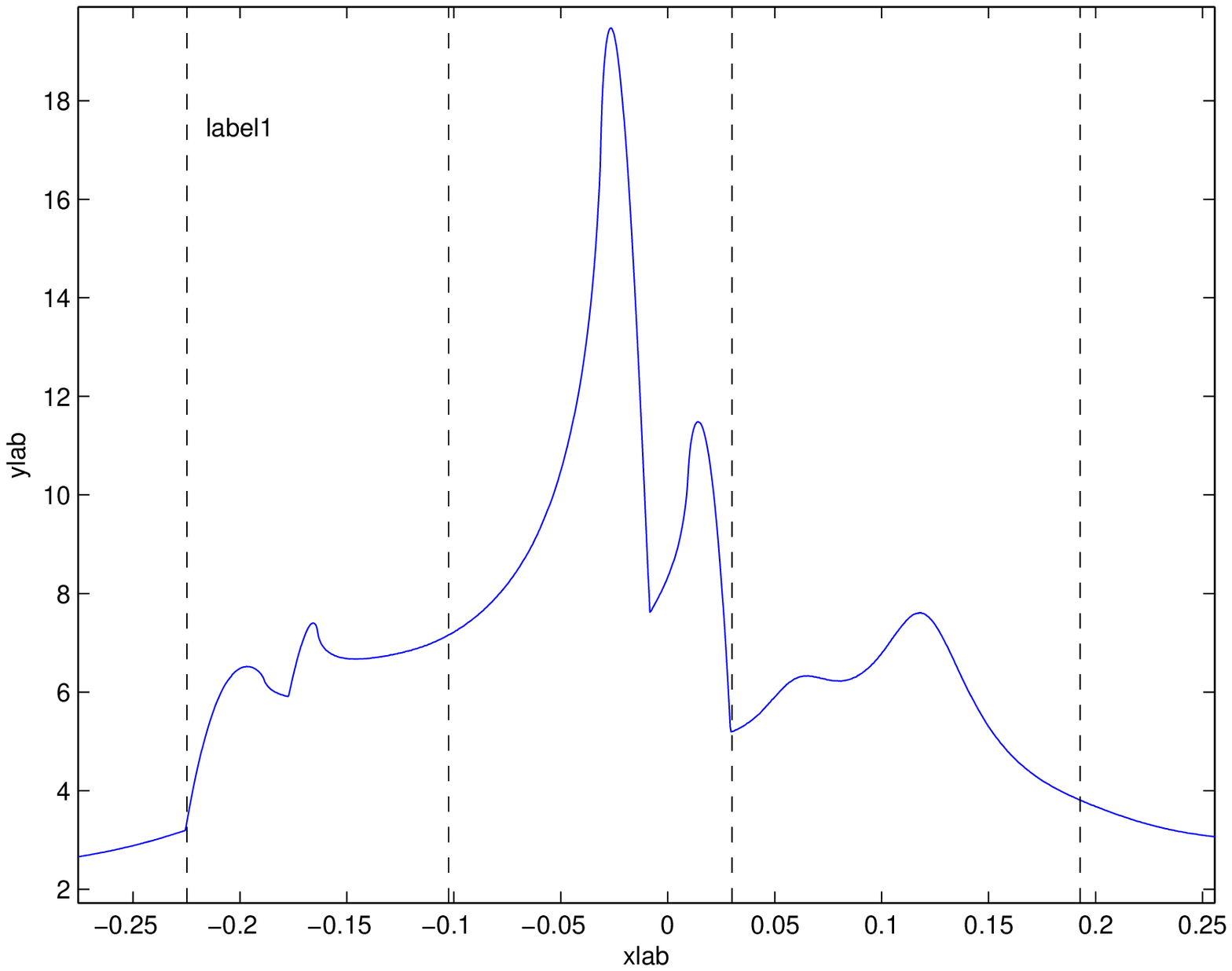}} 
\centering\includegraphics[width=0.45\hsize]{\FigDir{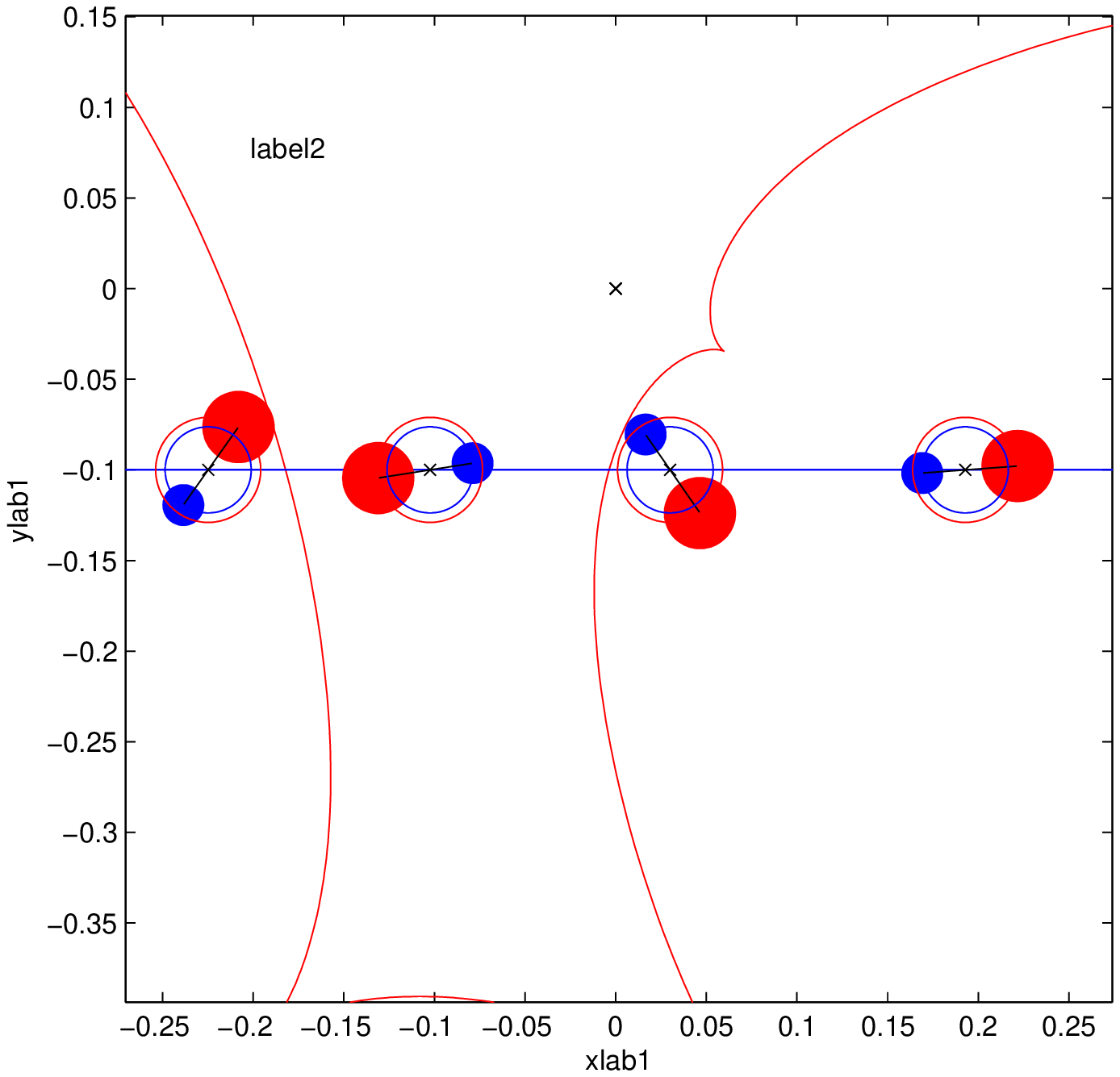}}\\
\centering\includegraphics[width=0.45\hsize]{\FigDir{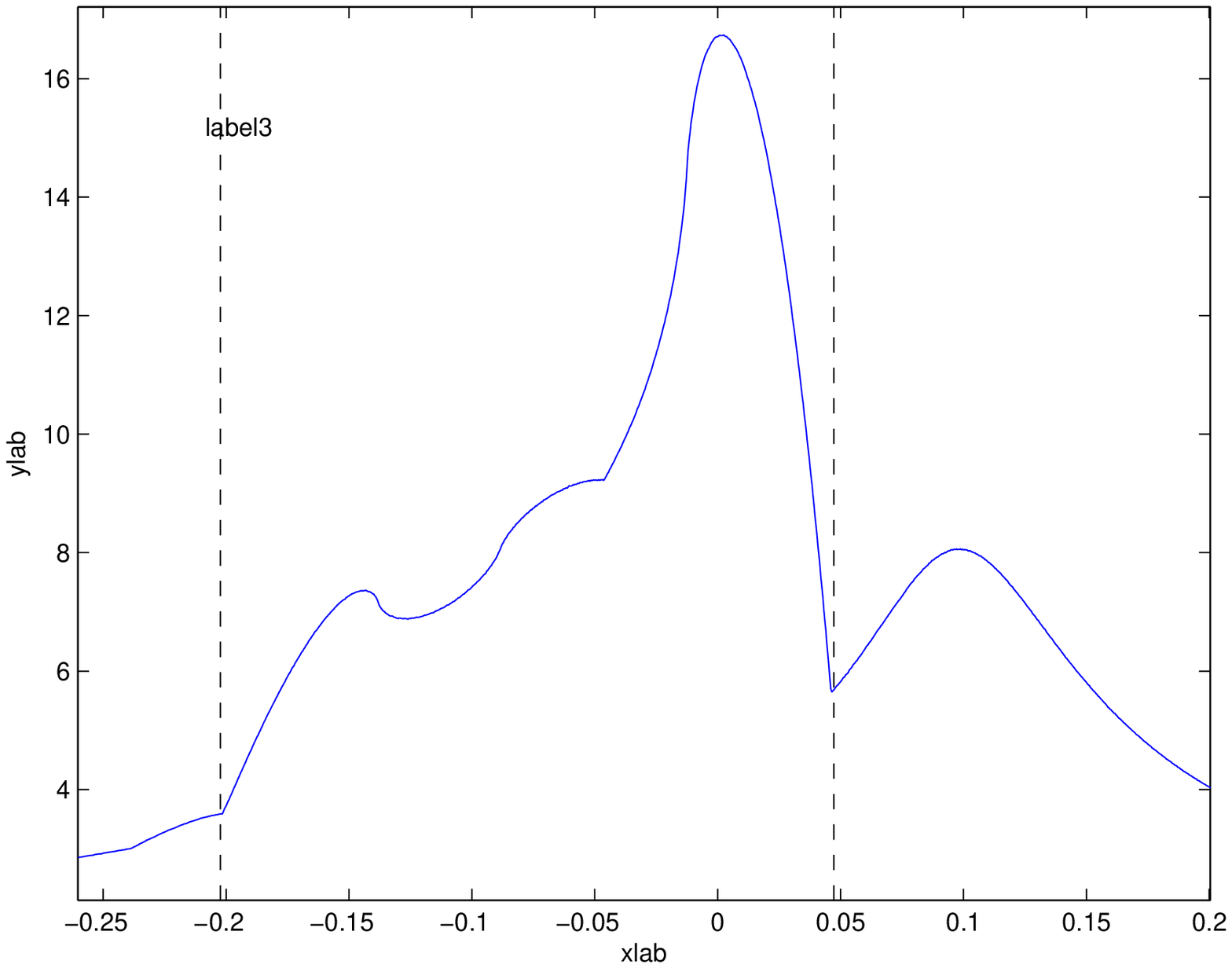}}
\centering\includegraphics[width=0.45\hsize]{\FigDir{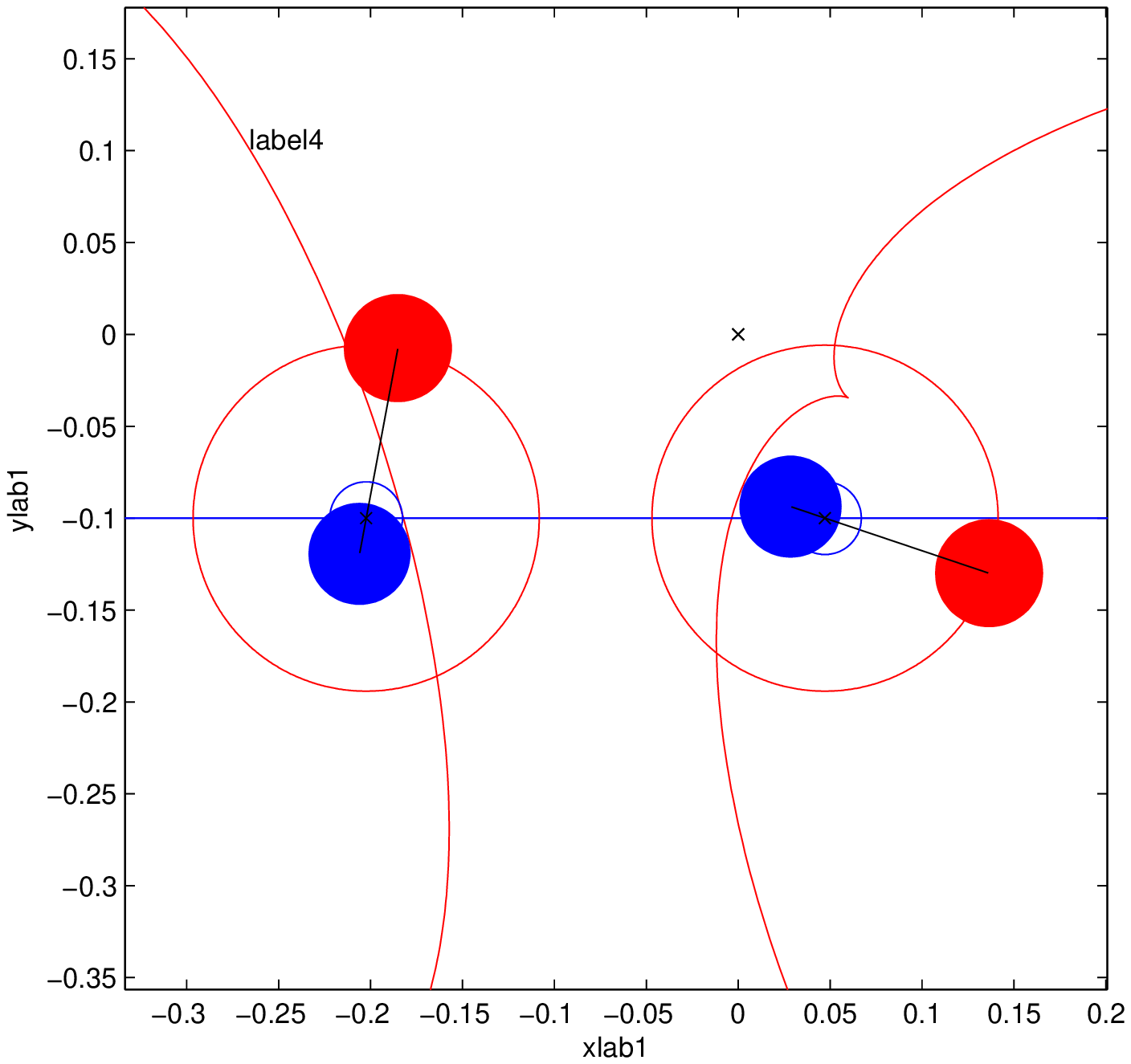}}\\
\centering\includegraphics[width=0.45\hsize]{\FigDir{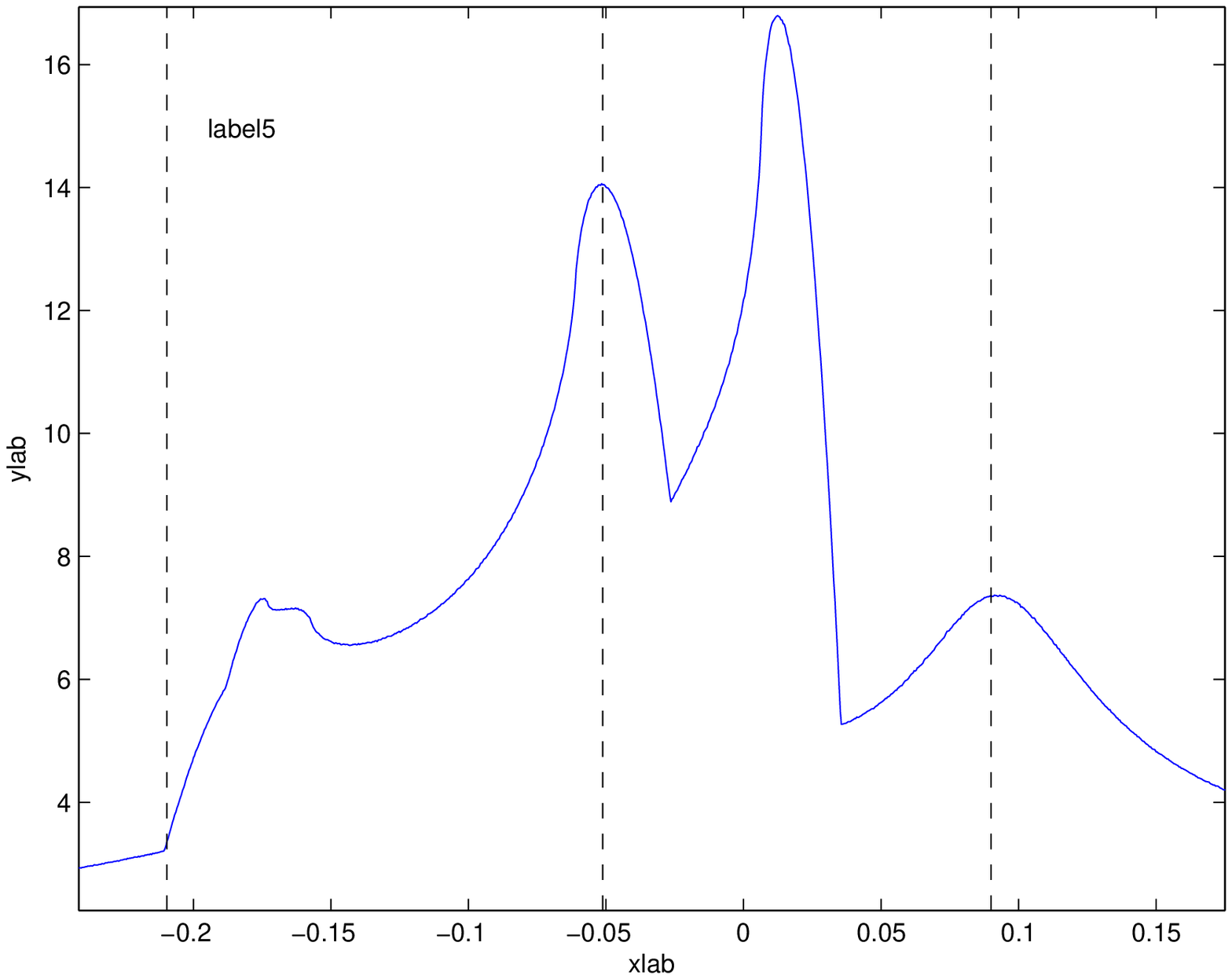}}
\centering\includegraphics[width=0.45\hsize]{\FigDir{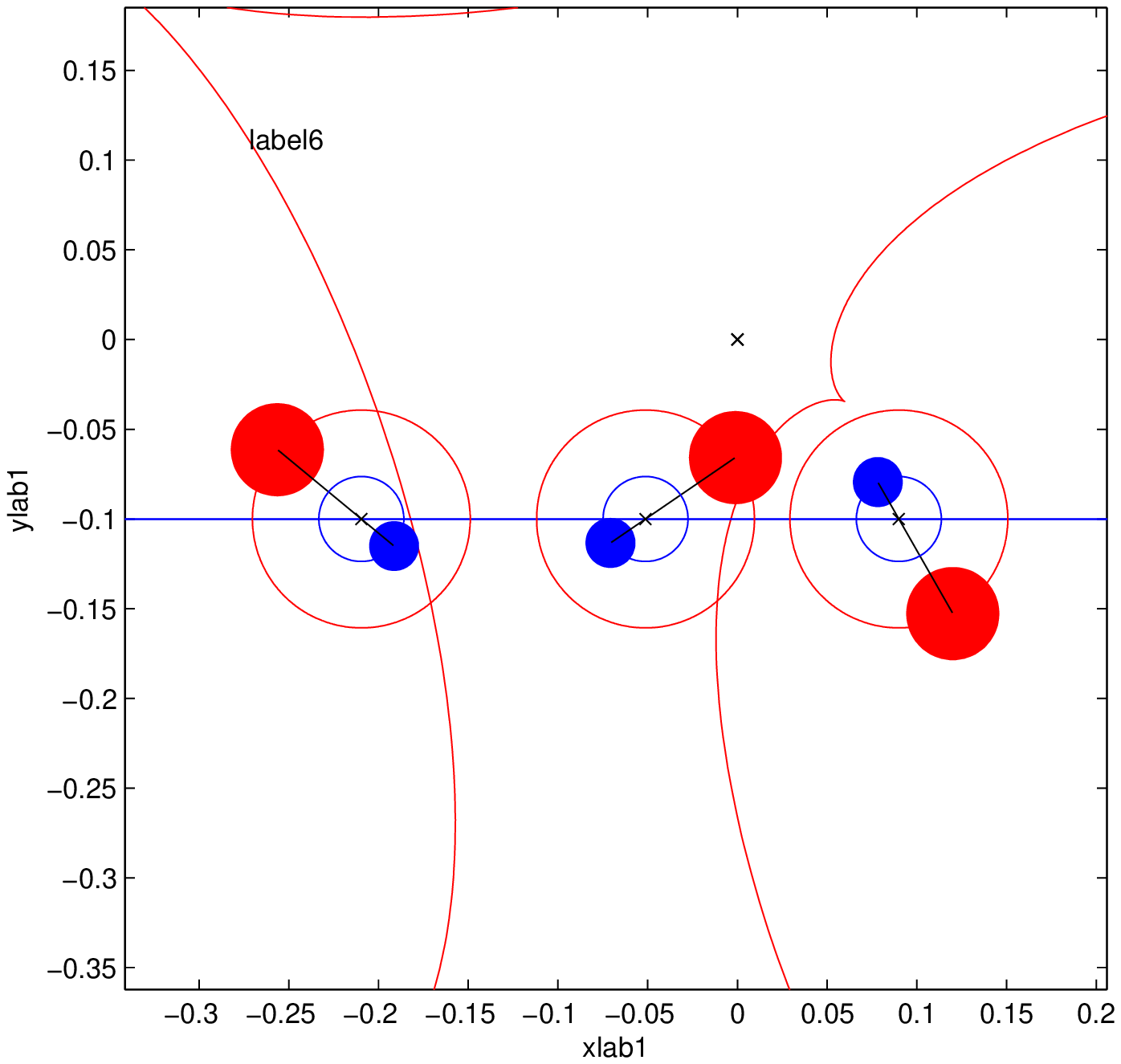}}
\caption{Example face on close binary star systems lensed by the binary lens described in the caption to Figure~\ref{fig:TWAnd}. Light curves for the close binary systems HH Carinae (top), RZ Scuti (middle) and V356 Sagittarii (bottom) assuming a face-on orientation ($i=0$) are shown on the left with the corresponding source system track and caustic plots shown on the right. The source system parameters are listed in Table~\ref{tab:eclipdata}. The source systems showing the relative stellar radii and change in orbital phase corresponding to the real stellar systems are shown at several times and indicated by vertical lines in the light curve plots. In each case the centre-of-mass of the binary source system moves along the source system track, shown in blue in the caustic diagrams, from left to right.}
\label{fig:matrix}
\end{figure*}

 The light curves and corresponding caustic line plots for the close binary star systems HH Carinae, RZ Scuti and V356 Sagittarii are shown in Figure~\ref{fig:matrix}; in each case with the real inclination angle changed from $i\simeq\pi/2$ to $i=0$, to create a non-eclipsing binary star source. The binary source parameters for these stars are listed in Table~\ref{tab:eclipdata}. In several of the light curves in Figure~\ref{fig:matrix} we see a repetition of a caustic crossing, in others, we see similar light curve features (e.g. Figure~\ref{fig:matrix}a) where the source stars pass close to a caustic line cusp. Other light curves show extended plateaux  where the orbital motion of the source star system conspires to keep at least one star close to a caustic cusp. The light curve features shown in  Figure~\ref{fig:matrix}, particularly the apparent repetition of a caustic crossing would be characteristic of a close binary system acting as the source and would be difficult to reproduce by invoking other physical phenomena.

Caustic crossing microlensing events may therefore be a discovery channel of close binary systems. The source radii of stars in an eclipsing binary system are measured routinely. The radii of non-eclipsing binary stars may be estimated in caustic crossing events through modelling not greatly dissimilar to that which is routinely carried out for singular source star events amplified by binary lenses. Estimates on source star radius in microlensing events have uncertainties ranging from $\sim 7\%$ -- 20\% (see e.g. \citealt{1997ApJ...491..436A, 1999ApJ...522.1011A,2000ApJ...541..270A,2001ApJ...549..759A,2002ApJ...572..521A,2004ApJ...617.1307J,2004ApJ...603..139Y}). In comparison, the median uncertainty on the fitted primary/large (secondary/small) star radii (as a fraction of orbital semi-major axis) in the catalogue of eclipsing binaries of \citet{2005ApJ...628..411D} is 9\% (17\%).

\clearpage

\clearpage
\subsection{Extended, non-circular source stars}

\begin{figure}
\begin{center}
\psfrag{xlab1}[][]{$\tN$}
\psfrag{ylab1}[][]{Amplification}
\psfrag{xlab2}[][]{$\rE (\times 10^{3})$}
\psfrag{ylab2}[][]{$\rE (\times 10^{3})$}
\psfrag{xlab3}[][]{$\rE$}
\psfrag{ylab3}[][]{$\rE$}
\hspace{-1cm}
\centering\includegraphics[width=1.0\hsize]{\FigDir{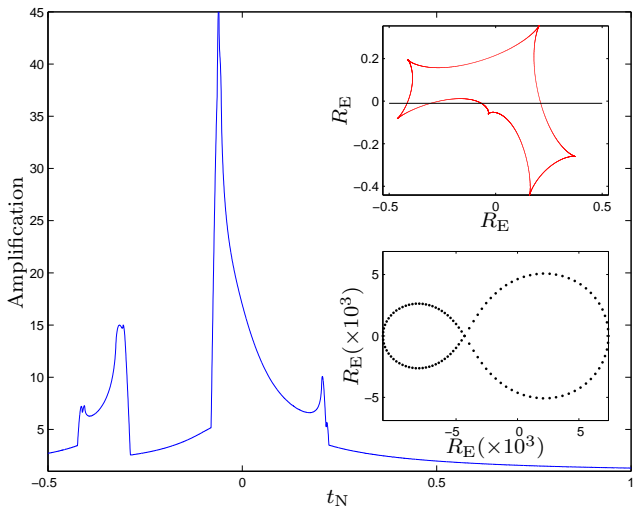}}
\end{center}
\caption{Light curve assuming a binary lens system with $\umin=0.01$, $q=0.11$, $d=0.95$, $\beta=120^{\circ}$. The caustic lines of this lens system with the source star system track is shown in the upper inset. The profile of the close binary system DM Delphini was used as the  source. The lower inset shows the stellar atmosphere profile of DM Delphini assuming the Roche model and stellar parameters from \citet{1987A&AS...67...87G}, but assuming $i=0$.}
\label{fig:DMDel_roche}
\end{figure}

\begin{figure*}
\psfrag{xlab1}[][]{\tN}
\psfrag{ylab1}[][]{Amplification}
\psfrag{xlab}[][]{$\rE$}
\psfrag{ylab}[][]{$\rE$}
\psfrag{label1}[][]{a}
\psfrag{label2}[][]{b}
\psfrag{label3}[][]{c}
\psfrag{label4}[][]{d}
\centering\includegraphics[width=0.45\hsize]{\FigDir{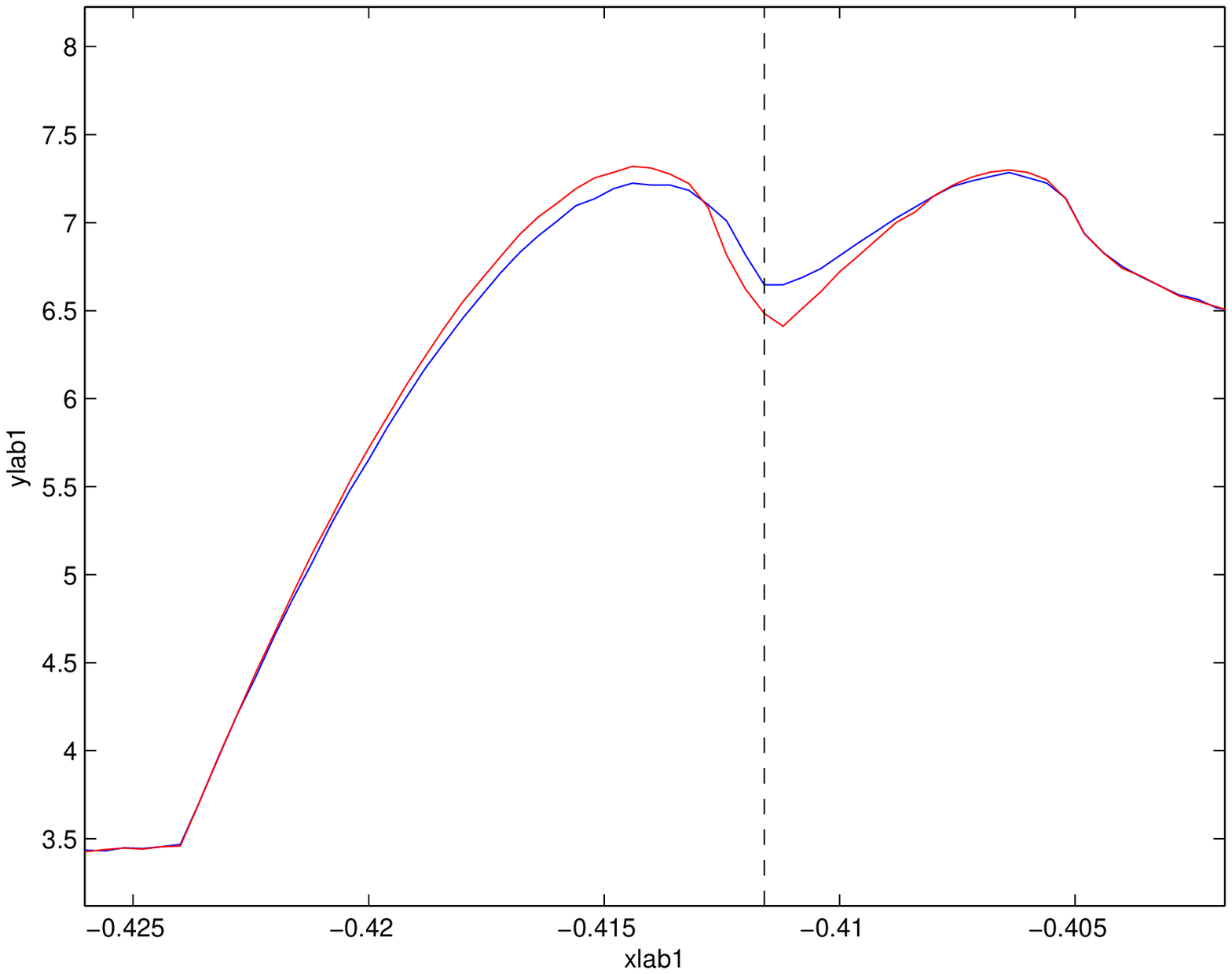}} 
\centering\includegraphics[width=0.45\hsize]{\FigDir{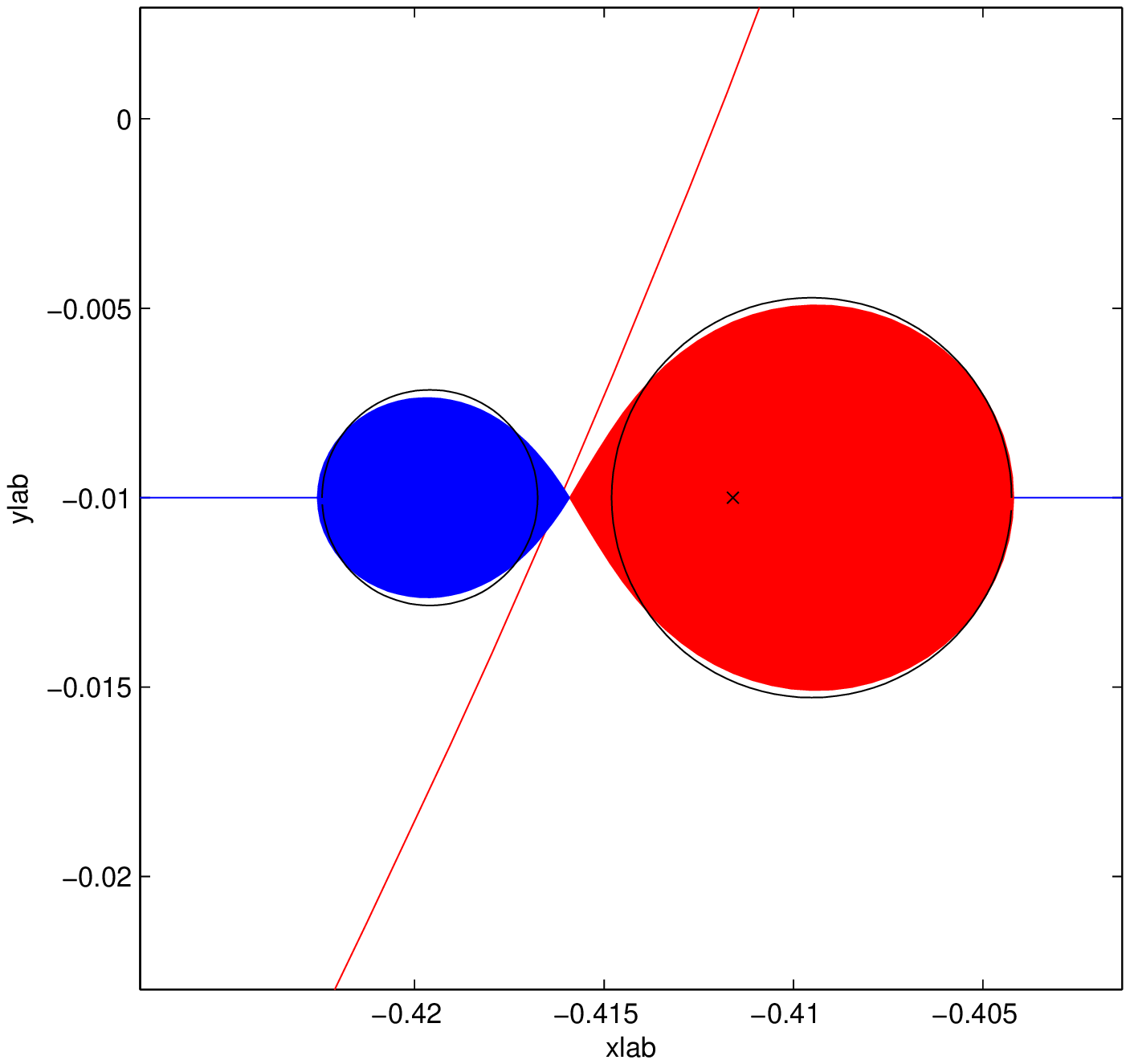}}\\
\centering\includegraphics[width=0.45\hsize]{\FigDir{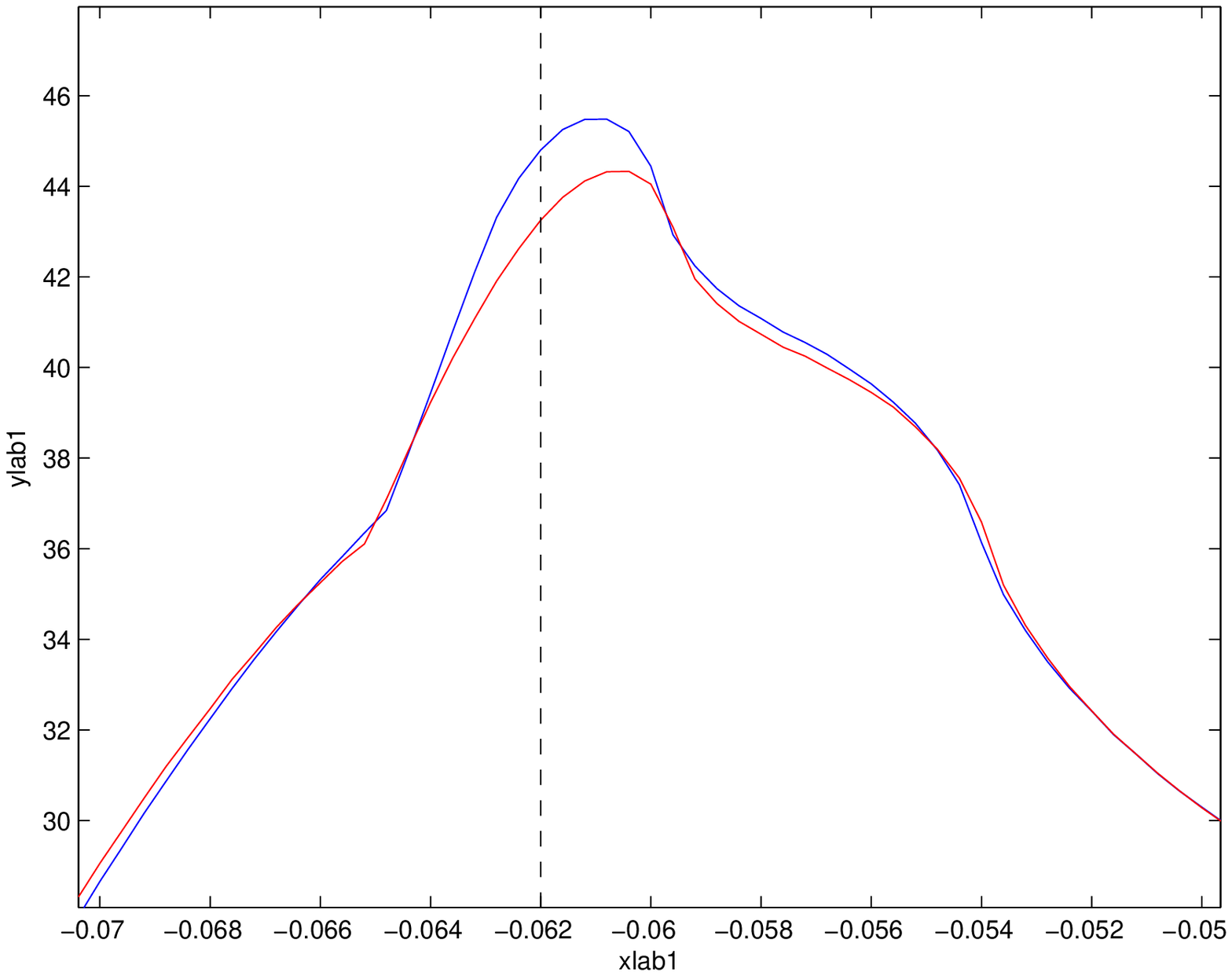}} 
\centering\includegraphics[width=0.45\hsize]{\FigDir{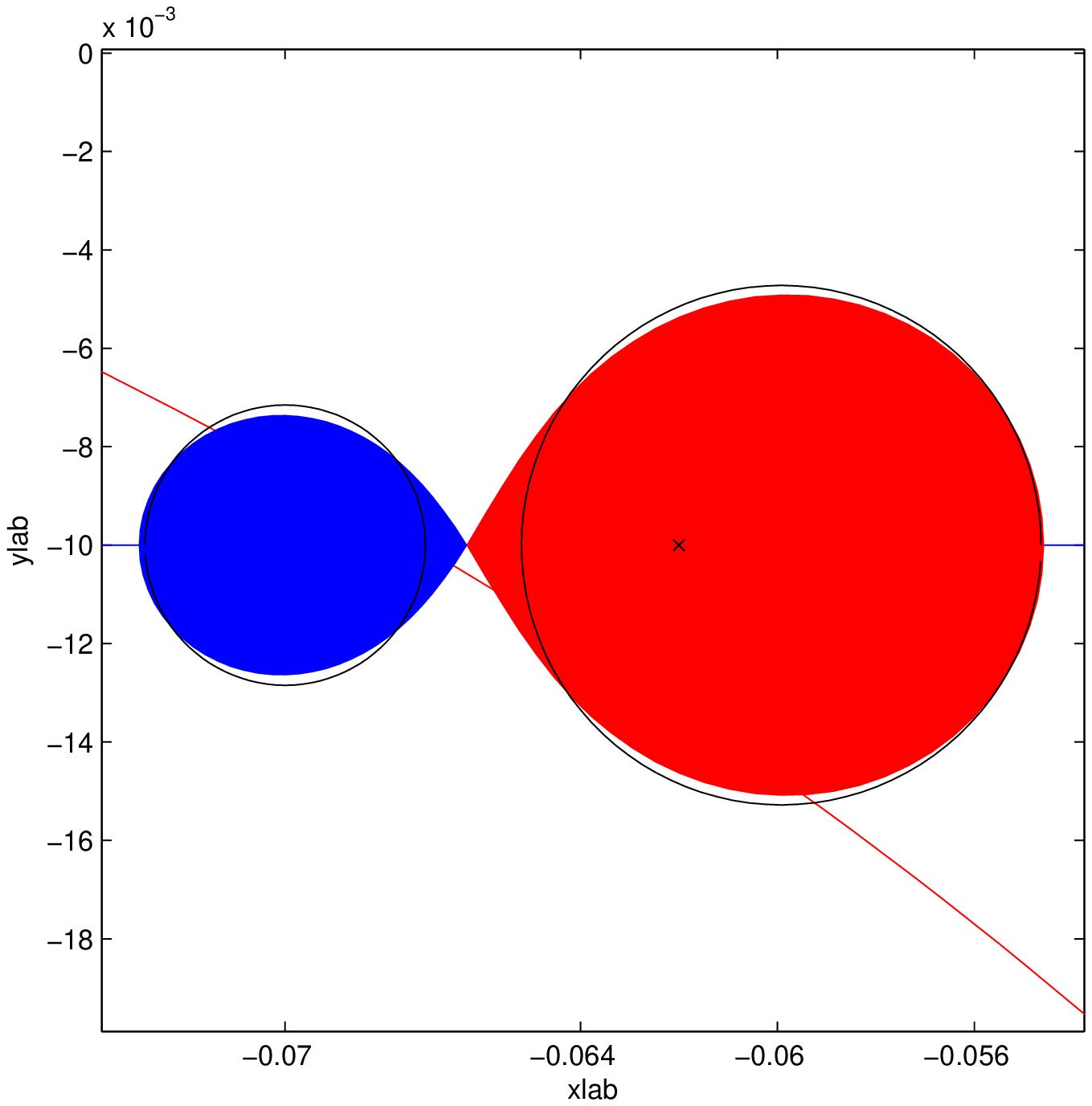}}\\
\caption{Light curves (left) and corresponding caustic crossing plots (right) assuming a binary lens system as described in the caption to Figure~\ref{fig:DMDel_roche} and a close binary star modelled on DM Delphini with inclination angle $i=0$. The light curve corresponding to the profile of the distended stellar atmosphere is shown in blue, compared to that produced by circular stellar profiles, which is shown in red. The face-on profile of DM Del is shown compared to circular profiles which contain the same area as the distended profiles. The two sets of axes show a caustic crossing where the tangent to the caustic is approximately perpendicular (top) and parallel (bottom) to the line of source star centres.}
\label{fig:matrixDMDel}
\end{figure*}

We consider now whether the spatial resolution afforded by the extreme amplification which occurs when a source star intersects a caustic line is sufficient to measure any deviation of the source star(s) from spheroidal. Close binary star systems are routinely modelled assuming that one or both of the stars' atmospheres have become distended due to the gravitational field of its companion. As an example, we compute the three dimensional shape of the binary star DM Delphini using the Roche model (e.g. \citealt{2001icbs.book.....H}) and stellar and orbital parameters as given in \cite{1987A&AS...67...87G}. The binary star envelope is shown in the lower inset of Figure~\ref{fig:DMDel_roche}, assuming an orbital inclination $i=0$. We take the two dimensional envelope of the full 3D solution of the binary star surfaces as the source profile for a caustic crossing microlensing event. We have assumed that both components of the binary source have the same surface brightness for simplicity. Figure~\ref{fig:DMDel_roche} shows the light curve for a caustic crossing event and Figure~\ref{fig:matrixDMDel} shows details of the light curves at times of two caustic crossings. The light curve for the same system using circular source star profiles is also shown for comparison purposes. The radius of each circular source star's profile is equal to the distance to each star's distended envelope in the direction directly opposite its companion. The total area of each circular profile was made to be the same for the circular and distended cases. This is important as caustic crossing features tend to be washed out to a greater degree as the source star size increases (see Figure~\ref{fig:geometry2}). For simplicity, we have neglected binary source orbital motion.

From  Figure~\ref{fig:matrixDMDel} it is clear that there is a difference in the shape of the light curves during a caustic crossing for the distended source profiles compared to circular sources. There is a clear excess of light in the light curves during the period between the source stars crossing the same caustic for the distended profiles. The appearance of clear ``repeated'' caustic crossing peaks will occur for events where the tangent to the caustic is perpendicular to the line connecting the source star centres. For events where the caustic line is more parallel to the binary star positional angle in its orbital plane as in the lower pair of axes in Figure~\ref{fig:matrixDMDel}, the separate light curve peaks merge. While the difference between the light curves assuming circular and distended source stars are clear in this comparison, it is likely that such a feature due to distended source stars could be modelled instead by altering (circular) source star radii. We return to this issue in Section~\ref{sec:conclusion}.

\subsection{Single lens microlensing}
\label{sec:single}
\begin{figure*}
\begin{center}
\psfrag{xlab1}[][]{$\tN$}
\psfrag{ylab1}[][]{Amplification}
\psfrag{xlab2}[][]{$\rE$}
\psfrag{ylab2}[][]{$\rE$}
\centering\includegraphics[width=0.45\hsize]{\FigDir{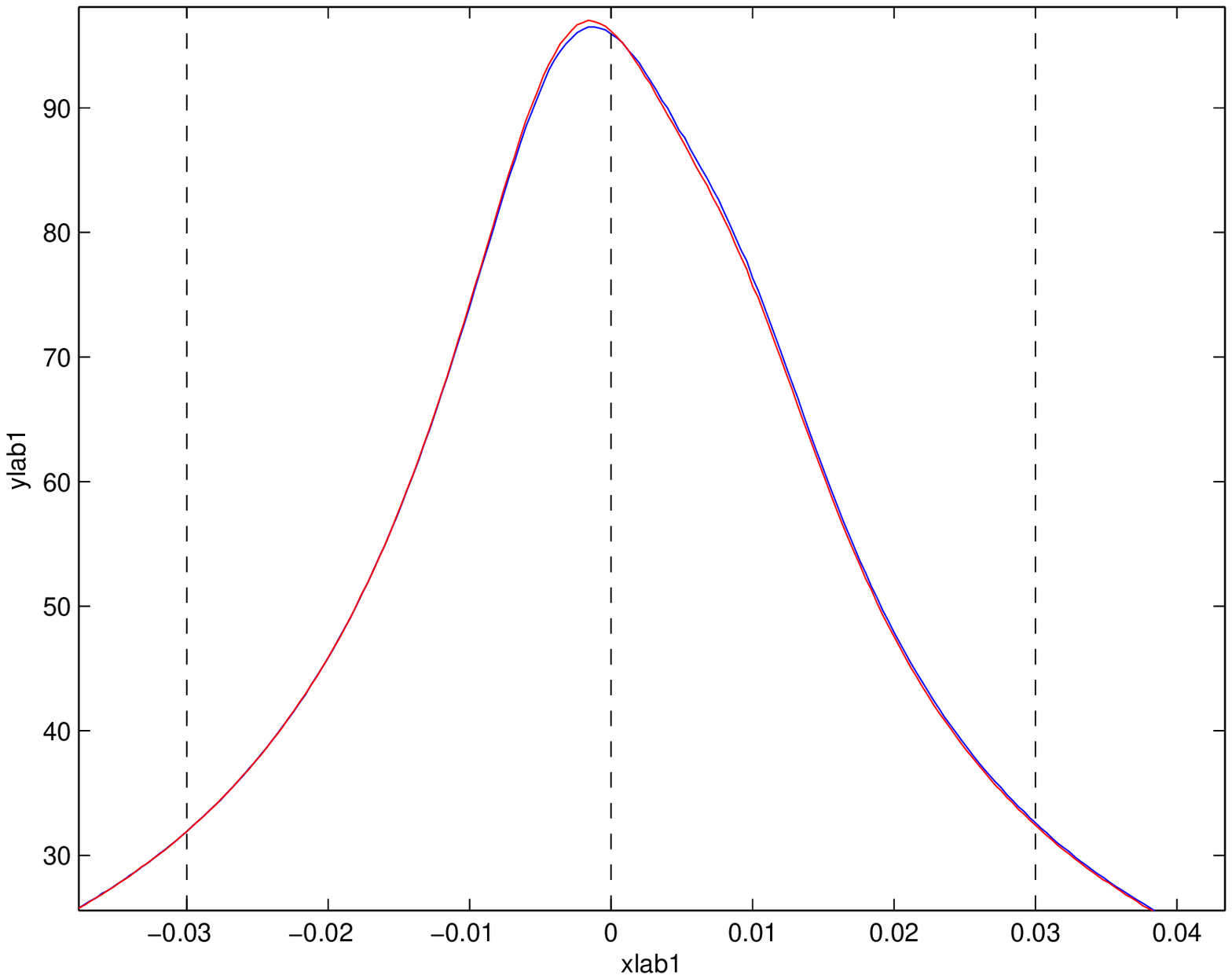}}
\centering\includegraphics[width=0.45\hsize]{\FigDir{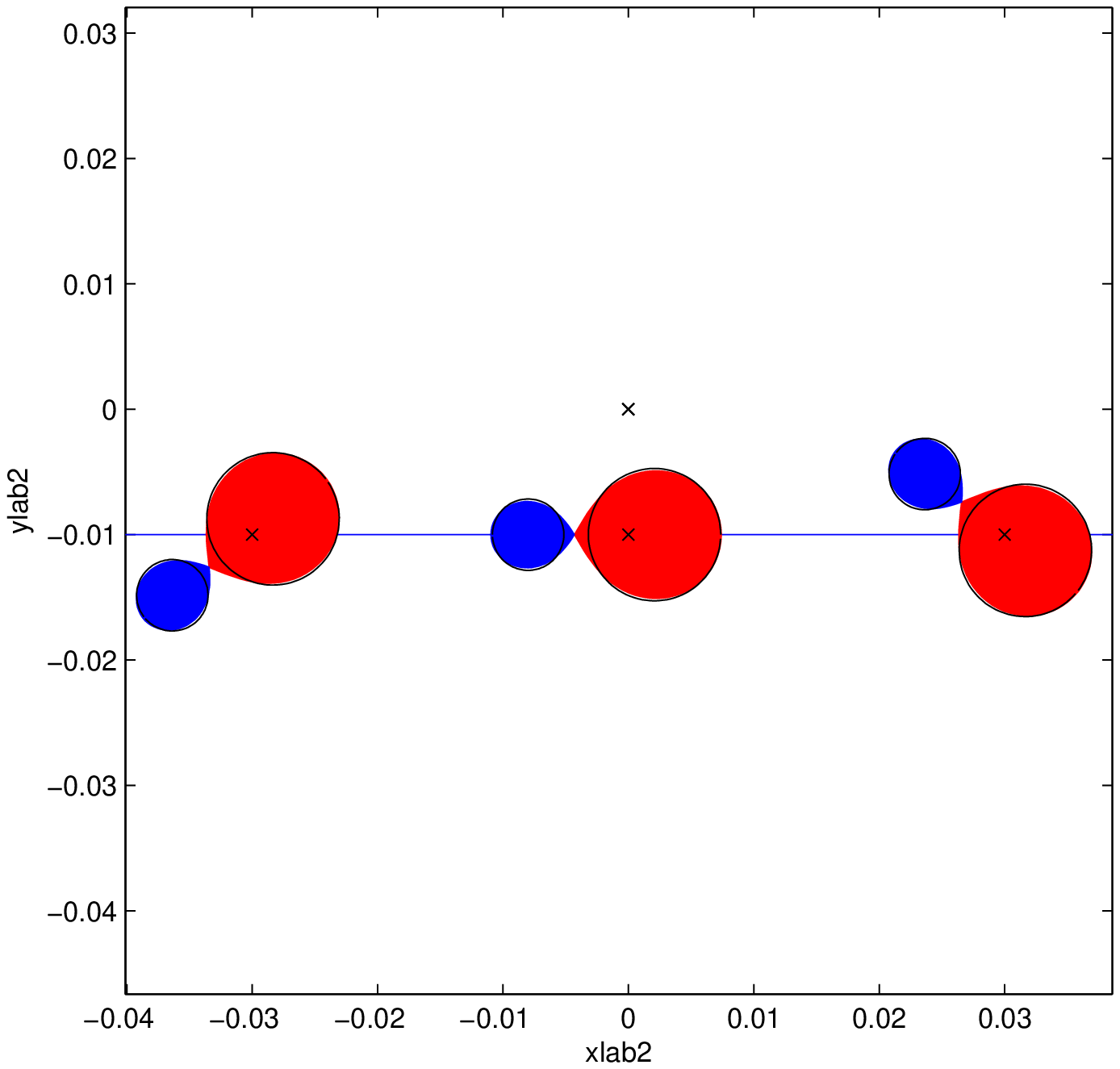}}
\end{center}
\caption{Light curves (left) and corresponding source system plot (right) assuming a single mass lens system with minimum impact parameter $\umin=0.01$. The light curve generated assuming a close binary source star modelled on DM Delphini with inclination angle $i=0$ is shown in blue, and compared to that produced assuming spherical source stars (red). Source orbital motion is included, and the position of the source system at three epochs is shown in the right plot, corresponding to the vertical dashed lines in the light curve plot on the left. The circular source profiles (shown in black) encompass the same area as the non-spherical source profiles. The source system centre-of-mass moves along the horizontal blue line from left to right.}
\label{fig:DMDel_highmag}
\end{figure*}

Extreme amplification of a background source can be achieved with a single lensing object, if the minimum impact parameter is sufficiently small $\umin \lesssim 0.02$. Several such extreme events have been observed, (see e.g. \citealt{Phil08,2006ApJ...642..842D}). The light curves arising assuming the distended source profile of the close binary system DM Del were compared to those generated assuming spherical stars for a high amplification, single lens event with $\umin = 0.01$. The light curve and source star track plot are shown in Figure~\ref{fig:DMDel_highmag}. In this case, the orbital motion of the binary source star was included, and can be seen as an asymmetry in the light curve around the time of peak amplification. The difference between the light curves assuming non-circular and circular star source profiles is however, small in comparison to deviations seen above for caustic-crossing events. Given an observed light curve which shows periodic amplitude fluctuations suggesting a close binary source, it might be possible nevertheless to include non-circular source profiles in the modelling of such curves. Conclusions on any departure from circular source profiles in such events would however require exquisite light curve data. We therefore restrict the discussion in Section~\ref{sec:discussion} to caustic crossing events.

\section{Discussion}
\label{sec:discussion}

Binary (or indeed multiple) element lens systems can produce remarkably diverse light curves even for a single source star. By adding a second source star these light curves naturally increase in complexity. After many years' operation, the microlensing survey collaborations are beginning to compile light curves which do not submit to standard analyses. There are many phenomena which could affect microlensing light curves, such as the Earth's orbital motion (parallax: see e.g. \citealt{2007ApJ...664..862D}) or that of a large radius source binary (xallarap: see e.g. \citealt{2005ApJ...633..914P}) or the rotation of the binary lens. However some observed light curves remain that show features that cannot be explained by invoking these physical interpretations. We also expect, from population statistics, a certain number of close binary stars to act as the source in microlensing events. In order to obtain as much information on these systems as possible, it is necessary to ensure that the survey and follow-up collaborations are aware of the possibility of close binary source microlensing.

The OGLE and MOA collaborations currently perform microlensing surveys of many fields toward the Galactic bulge and the Magellanic clouds. The OGLE collaboration \citep{2000AcA....50....1U} is currently operating its third evolution of its experiment, OGLE-III, surveying $\simeq 96$ square degrees toward the Galactic centre using the 1.3m Warsaw telescope at La Silla, Chile. The MOA collaboration observes $\sim 50$ square degrees toward the GC using the 1.8 MOA-II telescope in the South Island of New Zealand \citep{2001MNRAS.327..868B, 2003ApJ...591..204S}. Both collaborations monitor millions of stars per night and report the discovery of approximately 1000 microlensing events per year. 

In order to make predictions on the number of microlensing events per year which may show evidence for a close binary source, we estimate the number of events, $N = \Nul \cdot \Pcb \cdot \Pbl \cdot \Pcc \cdot \Pgv$ where \Nul is the number of microlensing events per year, \Pcb is the probability of a close binary star as the source of a microlensing event, \Pbl is the probability of a binary lens, \Pcc is the probability of a caustic crossing and \Pgv is the probability of being able to see the effects of distended stellar atmospheres during the caustic crossing. 

\Pgv will depend on the binary source orbit position angle with respect to the tangent to the caustic line, taking into account the orbital inclination of the binary source system. It can be shown from geometric arguments that a fair estimation of the probability of the interesting distended region between the two stars being optimally magnified by a caustic line is $\Pgv = 1 - 2\thc/\pi$ where \thc is the angle subtended at the internal similitude centre for the circular profiles corresponding to the radii of the binary star sources. The mean (median) value of \Pgv for all eclipsing binary stars in the catalogue of \citet{2004yCat.5115....0S} is 0.66 (0.67). 

The probability of a caustic crossing given a binary lens system, \Pcc, was estimated via a Monte Carlo analysis. $2\times10^{6}$ binary lens microlensing light curves were generated and \Pcc computed as the number of caustic crossings divided by the total number of lens systems tested, giving $\Pcc \simeq 0.1$.

We estimate the probability of a close binary system acting as the source in a microlensing event by considering the set of eclipsing binaries discovered in the OGLE-II database. \citet{2005ApJ...628..411D} found 10862 eclipsing binary stars in the OGLE-II database, which contains the light curves of $\sim 30\times 10^{6}$ stars, giving the fraction of observed eclipsing binaries as $\simeq 3.6\times10^{-4}$. As noted above, the effect of distorted source envelopes may be visible for non-eclipsing systems in caustic crossing events. We therefore consider the range of inclination angles of the eclipsing binary systems reported by \citet{2005ApJ...628..411D} and estimate that the total number of close binary systems is approximately a factor 4 greater than observed, giving $\Pcb \simeq 0.002$.

Assuming the binary lens fraction \Pbl = 0.5, we obtain $N \simeq 3\times10^{-4} \Nul$. Given the returns from current microlensing surveys of $\simeq 1000$ events per year, we would expect only one event over a five year period. The crude order of magnitude event rate estimate is subject to some serious assumptions. Foremost is the assumption that all close binary systems display distended atmosphere effects, which is clearly a gross over-estimation. We also implicitly assume that given the rare situation postulated here the light curve sampling rate, quality and coverage will be sufficient to trace the effects of distended stellar atmospheres. These requirements on the light curve data overlap those for detecting planets through microlensing, which has been successful. 
The low predicted event rate should not deter speculation on whether such systems will be observed in the future, either with current or future observational surveys. Nor should the predicted rarity of such events prevent some investigation into what analysis would be theoretically possible on the source star system in such events. The expected rate of a close binary star system acting as the source in a caustic crossing microlensing event, releasing the requirement on the source system to cross the caustic line in an optimal fashion to investigate any distention of the source atmospheres, is $\simeq 2$ events over a five year observation period. We therefore expect that a number $\mathcal{\rm few}$ events in the OGLE database to be due to close binary star source stars.

The observed event rate for close binary source star microlensing events is based on the current performance of the existing survey collaborations, OGLE and MOA. Proposals are been raised to champion a new paradigm of microlensing survey, whereby fewer fields are observed in the rich stellar fields of the Galactic bulge, but at maximum cadence allowed by the instrumentation. This ``Earth-Hunter'' scheme is predicted to return $\sim 6000$ microlensing events per year, almost a magnitude more than the current surveys. Consequently, we would expect a few close binary star source events per year. 

The first direct measurement of a star filling its Roche lobe was only recently achieved, using the Very Large Telescope Interferometer \citep{2007A&A...470L..21V}.  The extreme amplification afforded by binary lens systems where the shape of a close binary source star can be resolved in caustic crossing microlensing events may be one way these measurements can be made. Similarly, accretion disks and stellar hot spots may be resolvable in rare systems through gravitational lensing, an interesting topic outside the scope of this current work. 

The detection of extra-solar planets via microlensing has required in some cases the accurate modelling of light curve features which have a signal amplitude similar to the difference between the light curves assuming circular and non-circular source profiles in Figure~\ref{fig:matrixDMDel}, see for example \citet{2008arXiv0809.2997D,2008Sci...319..927G,2006ApJ...644L..37G,2006Natur.439..437B}. This would suggest that the current microlensing survey and follow-up networks of 0.5m -- 2m class telescopes (see \citealt{2006MPLA...21..919R} for a review) would produce data that could allow the discrimination between circular and non-circular source star profiles.

Light curves presented in this work assume the two stars in the close binary systems have the same emission spectra. In general this will not be the case. Multi-colour observations would be useful for modelling close binary source stars, as the additional colour information would enable a more accurate estimate for each star's radius. Multi-wavelength observations would also aid the discrimination between circular and non-circular source profiles, and enable a more accurate description of other atmospheric effects such as limb-darkening.

\section{Conclusion}
\label{sec:conclusion}
The extreme amplification of background sources by the gravitational effect of binary stars has yielded spectacular astrophysical results over the past decade. One aspect of the high amplification afforded by so-called caustic-crossing microlensing events is the ability to spatially resolve features on the scale of the source star. We considered the possibility and likelihood of a close binary star acting as the source in a caustic crossing microlensing event and furthermore, whether the effect of distended stellar atmospheres could be appreciable in such events. The signature of close binary source star is likely to be immediately apparent in a light curve by the presence of ``repeated'' caustic crossing features. It is clear that some additional flux can be present due to the distended source star profiles, compared to those corresponding to spherical stars. For this reason, if any light curve shows features that suggest a close binary star source, the modelling of the light curve should explore the possibility of including non-circular source star profiles for the source system.

The probability of observing such events is optimistically estimated at a few events over several years' monitoring. A new survey paradigm such as the proposed Earth Hunter network of telescopes would increase the event rate to a few such events per year.

\section*{Acknowledgements}
The author thanks \L. Wyrzykowski, S. Mao and E. Kerins for their helpful discussions and comments on this paper.



\end{document}